\documentclass[a4paper,12pt]{article}

\usepackage{graphicx,amsmath,amsfonts,amssymb,cite}
\usepackage[dvips]{color}

\newcommand{\mj}{m_j}
\newcommand{\az}{\alpha Z}
\newcommand{\me}{m_e}
\newcommand{\mpr}{m_p}
\newcommand{\mf}{M_F}
\newcommand{\mi}{M_I}
\newcommand{\ga}{\gamma}

\newcommand{\eps}{\varepsilon}
\newcommand{\mub}{\mu_0}
\newcommand{\mun}{\mu_N}
\newcommand{\la}{\langle}
\newcommand{\ra}{\rangle}

\newcommand{\vecb}{\vec{B}}

\newcommand{\De}{\Delta}
\newcommand{\Ga}{\Gamma}

\newcommand{\tv}{\tilde{v}}
\newcommand{\tc}{\tilde{c}}
\newcommand{\om}{\omega}
\newcommand{\bv}{\breve{v}}
\newcommand{\bc}{\breve{c}}
\begin{document}
\begin{center}
\textbf{\large 
{$g$ factor of Li-like ions with nonzero nuclear spin}} 
\\ \vskip 1cm{} 
D.~L.~Moskovkin$^1$, V.~M.~Shabaev$^1$, 
and W.~Quint$^2$
\\ \vskip 0.1cm{}
\emph{${}^1$Department of Physics, St. Petersburg State
University, Oulianovskaya 1, Petrodvorets, 
\\
St. Petersburg
198504, Russia
\\ 
${}^2$Gesellschaft f\"{u}r Schwerionenforschung,
Planckstrasse 1, D-64291 Darmstadt, Germany}
\end{center}
\vskip 0.5cm{}

\begin{abstract}
The fully relativistic theory of the $g$ factor of Li-like ions
with nonzero nuclear spin is considered for the
$(1s)^2 2s$ state. The magnetic-dipole hyperfine-interaction
correction to the atomic $g$ factor is calculated including the 
one-electron contributions as well as the
interelectronic-interaction effects of order $1/Z$. 
This correction is combined with the
interelectronic-interaction, QED, nuclear recoil, and nuclear size 
corrections to obtain high-precision theoretical  
values for the $g$ factor of Li-like ions with nonzero nuclear spin.
The results can be used for a precise determination 
of nuclear magnetic moments from $g$ factor experiments.
\begin{flushright}
PACS number(s): 31.30.Jv, 31.30.Gs, 32.60.+i, 12.20.Ds
\end{flushright}
\end{abstract}

\section{\large Introduction}

Recent measurements of the $g$ factor of hydrogenlike carbon and
oxygen have reached an accuracy of about $2\cdot 10^{-9}$
\cite{her00,hae00,ver04}.
These experiments stimulated theoretical investigations of this effect
\cite{blu97,per97,bei00,cza01,kar01a,kar01b,sha01,mar01,gla02,
bei02,sha02a,yer02a,sha02b,yer02b,nef02,kar02,mosk04,mosk06}. 
Besides a new
possibility for tests of the  magnetic sector of quantum
electrodynamics (QED), these investigations have already provided
a new determination of the electron mass (see Refs. \cite{ver04,mota05} 
and references therein). Extensions of these measurements to systems
with higher nuclear charge number $Z$ and to ions with nonzero
nuclear spin would also provide the basis for new determinations
of the fine-structure constant \cite{kar01a,mota05,wer01,shagla06}, 
the nuclear magnetic moments \cite{wer01}, and the nuclear charge radii.
Investigations of the transition probability between the hyperfine-
structure components in hydrogenlike  bismuth 
\cite{sch94,sha98,sha02c,win99} clearly
indicated the importance of the QED correction to the $g$ factor
for  agreement between theory and experiment.

Extending theoretical description from H-like to Li-like
ions, one encounters a serious complication due to the presence of
additional electrons. A number of relativistic calculations of the 
$g$ factor of Li-like ions were carried out previously 
\cite{heg75,ves80,lin93, yan01, yan02, ind01}.
However, to reach the accuracy comparable to the one for
H-like ions, a systematic quantum electrodynamic (QED) treatment 
is required \cite{sha02b,shagla03,glasha04,glavol06,shaand06}. 
In the present paper, we consider the $g$ factor of lithiumlike
ions with nonzero nuclear spin. Calculations of the $g$ factor for
these ions should  include corrections depending on the nuclear
$g$ factor. Besides a simple lowest-order nuclear-spin-dependent
contribution, one should calculate the hyperfine-interaction
correction, including the interelectronic-interaction effects.
In the present paper, the magnetic-dipole hyperfine-structure 
correction is calculated in a wide range of the nuclear charge
number $Z=3-100$. The electric-quadrupole 
hyperfine-structure correction is evaluated as well. 
The calculations are
based on perturbation theory in the parameter $1/Z$. The
contributions of zeroth and first orders in $1/Z$ are taken into
account for the magnetic-dipole correction and the contribution of zeroth
order is taken into consideration for the electric-quadrupole one. 
The $1/Z$ correction is evaluated within a rigorous QED
approach. The obtained results are combined with the
other corrections to get accurate theoretical predictions for the
$g$ factor of lithiumlike ions with nonzero nuclear spin.

Experimental investigations in this direction are anticipated in
the near future at University of Mainz and GSI \cite{qui01}.

Relativistic units ($\hbar=c=1$) and the Heaviside charge unit 
($\alpha=e^{2}/4\pi,\,e<0$) are used in the paper. 
In some important cases, the final
formulas contain $\hbar$ and $c$ explicitly to be applicable for
arbitrary system of units.

\section{\large The $g$ factor in the lowest-order one-electron approximation}

We consider  a lithiumlike ion placed in a weak homogeneous
magnetic field $\vec{B}$ directed along the $z$ axis. Assuming
that the energy level shift (splitting) due to interaction of the
valent $2s$ electron with
$\vec{B}$ is much smaller than the hyperfine-structure splitting,
$\Delta E_{\rm mag}\ll \Delta E_{\rm HFS}$, we can express the
linear-dependent part of this  shift in terms of the $g$ factor,
\begin{equation}\label{magsplit}
\Delta E_{\rm mag}=g\mub B\mf ,
\end{equation}
where $\mf$ is the $z$ projection of the total atomic angular
momentum $F=j+I, j+I-1, ..., |j-I|$; $\mf=-F, -F+1, ..., F-1, F$;
$j$ and $I$ are the total electron and nuclear angular momenta,
respectively; $\mub=|e|/(2m_e c)$ is the Bohr magneton. To obtain $g$
in relativistic approximation to the lowest order, we have to
evaluate the expression
\begin{eqnarray} \label{eq2}
\De E_{\rm mag} = \langle
nljIF\mf|V_{\vecb}|nljIF\mf\rangle\,,
\end{eqnarray}
where
\begin{eqnarray} \label{Vtot}
V_{\vecb}=V_{\vecb}^{(e)}+V_{\vecb}^{(n)}\,,
\end{eqnarray}
\begin{equation}\label{Ve}
V_{\vecb}^{(e)}=-e(\vec{\alpha}\cdot
\vec{A})=\frac{|e|}{2}(\vec{\alpha}\cdot [\vecb\times\vec{r}])\,,
\end{equation}
the vector $\vec{\alpha}$ incorporates the Dirac $\alpha$
matrices, and
\begin{equation}\label{Vn}
V_{\vecb}^{(n)}=-(\vec{\mu}\cdot \vecb)\,.
\end{equation}
Here $V_{\vecb}^{(e)}$ describes the interaction of the valent electron
with the homogeneous magnetic field
and $V_{\vecb}^{(n)}$ describes the interaction of the nuclear magnetic moment
$\vec{\mu}$ with $\vecb$.
$|nljIF\mf\rangle $ is the atomic wave function that corresponds
to given values of $F$ and $\mf $. It is a linear combination of
products of electron and nuclear wave functions:
\begin{equation}\label{ket}
|nljIF\mf\rangle=\sum_{\mj, \mi}C_{j\mj
I\mi}^{F\mf}|nlj\mj\rangle|I\mi\rangle.
\end{equation}
Here $C_{j\mj I\mi}^{F\mf}$ are the Clebsch-Gordan coefficients;
$|nlj\mj\rangle$ are the unperturbed one-electron wave functions,
which are four-component eigenvectors of the Dirac equation for
the Coulomb field, with the total angular momentum $j$ and its
projection $\mj$; $n$ is the principal  quantum number
 and $l=j\pm \frac{1}{2}$ defines the parity of the
state. $|I\mi\rangle$ are the nuclear wave functions with the
total angular momentum $I$ and its projection $\mi$.

In what follows, we adopt for the nuclear magnetic moment $\mu
=\langle II|\mu{}_{z}|II\rangle$, where $\mu{}_z$ is the $z$
projection of the nuclear magnetic moment operator
 $\vec{\mu}$
acting in the space of nuclear wave functions $|I\mi\rangle$.

A simple integration over the angular variables in Eq. (\ref{eq2})
yields the well-known result (see, e.g., Ref. \cite{bet57})
\begin{equation}\label{gat}
g= g_{\rm D}\frac{F(F+1)+j(j+1)-I(I+1)}{2F(F+1)}-
\frac{\me}{\mpr}g_I\frac{F(F+1)+I(I+1)- j(j+1)}{2F(F+1)}\,.
\end{equation}
Here $m_e$ and $\mpr$ are the electron and the proton mass,
respectively, $g_{\rm D} $ is the one-electron Dirac value of the
ground-state $g$ factor of the lithium-like ion,  $g_I=\mu/(\mun I)$ 
is the nuclear
$g$ factor, and $\mun=|e|/(2\mpr c)$ is the nuclear magneton. To
obtain  $g_{\rm D} $ we have to calculate $\langle
nlj\mj|V_{\vecb}^{(e)}|nlj\mj\rangle $. For the point-nucleus
case, a simple evaluation for the $2s$ state yields
\begin{eqnarray}\label{gD}
g_{\rm D}=\frac{2[\sqrt{2+2\ga}+1]}{3} = 2 - \frac{(\az)^2}{6}+...\,,
\end{eqnarray}
where $\ga=\sqrt{1-(\az)^2}$.

 The $g_I$-dependent term in the right-hand side of equation (\ref{gat})
is by a factor ${\me}/{\mpr}\approx {1}/{1836}$ smaller than the
first term. However, since it is much larger than the current
experimental uncertainty, this equation can be used for
determination of $g_I$ from the $g$-factor experiments, provided
all the corrections to expression (\ref{gat}) are calculated to a
high accuracy or are known from the corresponding experiment for
another isotope of the same element. In particular, to meet the
level of the current experimental accuracy, we need to evaluate
the hyperfine-interaction correction.

\section{\large Hyperfine-interaction corrections}

The hyperfine-interaction operator is given by the sum
\begin{equation}
V_{\rm HFS}=V_{\rm HFS}^{(\mu)}+V_{\rm HFS}^{(Q)}\,,
\end{equation}
where $V_{\rm HFS}^{(\mu)}$ and $V_{\rm HFS}^{(Q)}$ are the
magnetic-dipole and electric-quadrupole hyperfine-interaction
operators, respectively. In the point-dipole approximation,
\begin{equation}\label{FB}
V_{\rm HFS}^{(\mu)}=
\frac{|e|}{4\pi}\frac{(\vec{\alpha}\cdot[\vec{\mu}\times
\vec{r}])}{r^3}\,,
\end{equation}
and, in the point-quadrupole approximation,
\begin{equation}
V_{\rm HFS}^{(Q)} = -\alpha \sum_{m=-2}^{m=2}Q_{2m}
\eta_{2m}^*(\vec{n})\,.
\end{equation}
Here $ Q_{2m}=\sum_{i=1}^Zr_i^2C_{2m}(\vec{n}_i)$ is the operator
of the electric-quadrupole moment of the nucleus,
$\eta_{2m}=C_{2m}(\vec{n})/r^3 $ is an operator that acts on
electron variables, $\vec{n}=\vec{r}/r$, 
$\vec{n}_i=\vec{r}_i/r_i$, $\vec{r}$ is the
position vector of the electron, $\vec{r}_i$ is the position
vector of the $i$-th proton in the nucleus, $C_{lm}=
\sqrt{4\pi/(2l+1)}\,Y_{lm}$, and $Y_{lm}$ is a spherical harmonic.
It must be stressed that the electric-quadrupole interaction
should be taken into account only for ions with $I > 1/2$.

In the one-electron approximation, the magnetic-dipole and 
electric-quadrupole hyperfine-interaction corrections 
to the ground-state $g$ factor of the Li-like ion are given by
\begin{equation}\label{HFScg}
\delta g_{\rm HFS(\mu, Q)}^{{\rm one-el.}(2s)}=\frac{2}{\mub B\mf}
\sum_{\mj\mi}\sum_{\mj'\mi'}C^{F\mf}_{\frac{1}{2}\mj I\mi}
C^{F\mf}_{\frac{1}{2}\mj' I\mi'}
\la I\mi|\sum_n^{(\eps_n\neq\eps_v)}
\frac{\la v|V_{\vecb}^{(e)}|n\ra\la n|V_{\rm
HFS}^{(\mu, Q)}|v'\ra}{\eps_v-\eps_n}|I\mi'\ra,
\end{equation}
where $|v\ra=|20\frac{1}{2}\mj\rangle$ and $|v'\ra=|20\frac{1}{2}\mj'\rangle$ 
are the $2s$ states of the valent electron with the
angular momentum projections $\mj$ and $\mj'$, respectively, 
$|n\ra\equiv |nlj\mj\ra$, $\eps_v=\eps_{2s}$ and $\eps_n$ 
are the one-electron Dirac energies in the Coulomb field of the nucleus.
The summation in (\ref{HFScg}) runs over
discrete as well as continuum states. 
The corresponding diagrams are presented in Fig. \ref{Diagrams2}.

The total hyperfine-interaction correction 
to the ground-state $g$ factor of the Li-like ion is given by
\begin{equation}\label{gcorrtot}
\delta g_{\rm HFS}^{(2s)}=\delta g_{\rm{HFS}(\mu)}^{(2s)}+ \delta
g_{\rm{HFS}(Q)}^{(2s)}
\end{equation}
with
\begin{equation}\label{gcorrdip}
\delta g_{\rm{HFS}(\mu)}^{(2s)}=\alpha^2
Z\frac{1}{12}\frac{\mu}{\mun}\frac{\me}{\mpr} \frac{1}{I}Y_{\rm
nuc}^{(\mu)}(F)[ S_2(\alpha Z) + \frac{1}{Z} B_{\mu}(\az) + 
\frac{1}{Z^2} C_{\mu}(\az) + ...]
\end{equation} 
and
\begin{equation}\label{gcorrquadr}
\delta g_{\rm{HFS}(Q)}^{(2s)}=\alpha^4 Z^3\frac{23}{2160}
Q\left(\frac{\me c}{\hbar}\right)^2 Y_{\rm nuc}^{(Q)}(F) 
[T_2(\alpha Z) + \frac{1}{Z} B_{Q}(\az)+ 
\frac{1}{Z^2} C_{Q}(\az)+ ... ]\,.
\end{equation}
Here the angular factors are
\begin{eqnarray} \label{Ynucdip}
Y_{\rm nuc}^{(\mu)}(F)=\frac{F(F+1)+I(I+1)-3/4}{2F(F+1)}=
\begin{cases}
  \frac{2(I+1)}{2I+1}   &\text{for $F=I-\frac{1}{2}$}\\
  \frac{2I}{2I+1}        &\text{for $F=I+\frac{1}{2}$}
\end{cases}
\end{eqnarray}
and
\begin{eqnarray} \label{Ynucquadr}
Y_{\rm nuc}^{(Q)}(F)=
\begin{cases}
  -\frac{(I+1)(2I+3)}{I(2I-1)(2I+1)}   &\text{for $F=I-\frac{1}{2}$}\\
  \,\,\,\,\,\,    \frac{1}{2I+1}        &\text{for $F=I+\frac{1}{2}$}
\end{cases}\,,
\end{eqnarray}
and $Q=2\langle II|Q_{20}|II \rangle$ is the electric-quadrupole
moment of the nucleus.
The functions
\begin{equation}\label{defS2}
S_2(\az)=\frac{12}{\alpha^2 Z\,\frac{\me}{\mpr} g_I
Y_{\rm nuc}^{(\mu)}(F)}\,\delta g_{\rm HFS(\mu)}^{{\rm one-el.}(2s)}
\end{equation}
and
\begin{equation}\label{defT2}
T_2(\az)=\frac{2160}{23 \alpha^4 Z^3\, Q
\left(\frac{\me c}{\hbar}\right)^2
Y_{\rm nuc}^{(Q)}(F)}\,\delta g_{\rm HFS(Q)}^{{\rm one-el.}(2s)}
\end{equation}
determine the one-electron contributions, which are discussed in 
detail in Ref. \cite{mosk04}.
For the point-charge nucleus, the functions $S_2(\az)$ and $T_2(\az)$ 
are~\cite{mosk04}
\begin{align}\label{S2}
S_2(\az)&=\frac{8}{3N}
\biggl\{\frac{1}{N+2}\biggl[N + \frac{10(N+1)}{3N}\biggr]
+\frac{(\az)^2}{\ga(\ga+1)}\biggl[\frac{2(N+1)}{3-4(\az)^2} + 1\biggr]
-\frac{1}{\ga}\biggr\}
\notag \\
&=1+\frac{229}{144}(\az)^2+...
\end{align}
and
\begin{align}
T_2(\az)&=\frac{192[(N+\ga+1) \{18 + 24\ga - 12N + 8\ga
N^2\} + 15(1+\ga)]} {23\ga N^3[15 - 16(\az)^2](N + \ga+1)^2} 
\notag \\
&=1+\frac{427}{276}(\az)^2+...\,,
\end{align}
where 
$N=\sqrt{2(1+\ga)}$.

The interelectronic-interaction correction $B_\mu(\az)$ is calculated 
within the rigorous QED approach. The interaction of the electrons 
with the Coulomb field of the nucleus is included in the unperturbed
Hamiltonian, i.e. the Furry picture is used. The perturbation theory 
is formulated with the technique of the two-time Green function (TTGF) 
\cite{sha02c}. To simplify the calculations, the closed
${(1s)^2}$ shell is regarded as belonging to a redefined vacuum. 
With this vacuum, the Fourier transform of TTGF can be introduced by
\begin{equation}
\begin{split}
{\cal{G}}(E;{\vec{x'}};{\vec{x}})&\delta(E-E')=\frac{1}{2\pi
i}\int\limits_{-\infty}\limits^{\infty}dx^{0}dx'^{0}
\exp(iE'x'^{0}-iEx^{0}) \\&
\times\langle0_{(1s)^2}|T\psi(x'^{0},{\vec{x'}})\psi^{\dag}(x^{0},
{\vec{x}})|0_{(1s)^2}\rangle,
\end{split}
\end{equation}
where $\psi(x^{0},{\vec{x}})$ is the electron-positron field
operator in the Heisenberg representation and $T$ is the
time-ordered product operator. The energy shift of a state $a$ can be 
expressed in terms of the TTGF defined by
\begin{equation}
g_{aa}(E) = \langle u_{a}|{\cal{G}}(E)| u_{a}\rangle \equiv \int
{d\vec{x}}d{\vec{x'}} u_{a}^{\dag}({\vec{x'}}){\cal{G}}(E;{\vec
{x'}};{\vec{x}}) u_{a}({\vec{x}}),
\end{equation}
where $u_{a}({\vec{x}})$ is the unperturbed Dirac wave function of the
state $a$.
Using the Sz.-Nagy and Kato technique \cite{SNK}, one can derive for the 
total energy shift $\De E_a\equiv E_a - E_a^{(0)}$ \cite{sha02c,MPEA}
\begin{equation}\label{DeEtot}
\De E_{a}=\frac{\frac{1}{2\pi i}\oint\limits_{\Gamma}{dE
\,\De E \De g_{aa}(E)}}
{1 + \frac{1}{2\pi i}\oint\limits_{\Gamma}{dE
\, \De g_{aa}(E)}}\,,
\end{equation}
where $\De E=E-E_a^{(0)}$, 
$\De g_{aa}(E)\equiv  g_{aa}(E)- g^{(0)}_{aa}(E)$, and 
$g^{(0)}_{aa}(E)=(E-E_a^{(0)})^{-1}$. The contour integrals in the
complex $E$-plane are taken along the contour $\Ga$ which
surrounds the pole of $g_{aa}(E)$ corresponding to the 
level $a$ and keeps outside 
all other singularities. The contour $\Gamma$ is oriented
counter-clockwise.

To first three orders of the perturbation theory, the energy
shift is given by
\begin{align} 
\De E_{a}^{(1)}&=\frac{1}{2\pi
i}\oint_{\Gamma}{dE \, \De E\De
g_{aa}^{(1)}(E)}\,,
\\
\label{DeE2}
\De E_{a}^{(2)}&=\frac{1}{2\pi
i}\oint_{\Gamma}{dE \, \De E\De
g_{aa}^{(2)}(E)}   
-\left(\frac{1}{2\pi i}\oint_{\Gamma}{dE \,
\De E \De g_{aa}^{(1)}(E)}\right) \left(\frac{1}{2\pi
i}\oint_{\Gamma}{dE \, \De g_{aa}^{(1)}(E)}\right) \,,
\\
\label{DeE3}
\De E_{a}^{(3)}&=\frac{1}{2\pi
i}\oint_{\Gamma}{dE \, \De E\De
g_{aa}^{(3)}(E)}   
-\left(\frac{1}{2\pi i}\oint_{\Gamma}{dE \,
\De E \De g_{aa}^{(2)}(E)}\right) \left(\frac{1}{2\pi
i}\oint_{\Gamma}{dE \, \De g_{aa}^{(1)}(E)}\right)
                      \nonumber \\
&-\left(\frac{1}{2\pi i}\oint_{\Gamma}{dE \,
\De E \De g_{aa}^{(1)}(E)}\right) \left(\frac{1}{2\pi
i}\oint_{\Gamma}{dE \, \De g_{aa}^{(2)}(E)}\right)
                     \nonumber \\
&+\left(\frac{1}{2\pi i}\oint_{\Gamma}{dE \,
\De E \De g_{aa}^{(1)}(E)}\right) \left(\frac{1}{2\pi
i}\oint_{\Gamma}{dE \, \De g_{aa}^{(1)}(E)}\right)^2 \,.
\end{align}

The redefinition of the vacuum changes $i0$ to $-i0$ in the
electron propagator denominators corresponding to the closed
$(1s)^2$ shell. In other words it means replacing the standard
Feynman contour of integration over the electron energy $C$ with a
new contour $C'$ (Fig. \ref{E-plane}). The second-order 
contribution is defined by the diagrams presented in Fig. 
\ref{Diagrams2}. Its evaluation according to Eq. (\ref{DeE2}) yields
formula (\ref{HFScg}). In the formalism under consideration, the
lowest-order interelectronic-interaction and the radiative
corrections to Eq. (\ref{HFScg}) are described by the third-order 
diagrams presented in Fig. \ref{Diagrams3} and, according to 
Eq. (\ref{DeE3}), by some products of the low-order diagrams
depicted in Figs. \ref{Diagrams31} and \ref{Diagrams32}. 
According to Fig. \ref{E-plane}, to separate 
the interelectronic-interaction corrections, the contour $C'$ must be 
divided into two parts, $C$ and $C_{\rm int}$.
The integral along the standard
Feynman contour $C$ gives the one-electron radiative correction.
The integral along the contour $C_{\rm int}$ describes the interaction
of the valent electron with the closed shell electrons. Its
evaluation in the Feynman gauge employing formula (\ref{DeE3})
yields for the interelectronic-interaction correction $B_{\mu}(\az)$ 
\begin{equation}\label{defBmu}
B_{\mu}(\az)=\frac{12}{\alpha^2\,\frac{\me}{\mpr}g_I
Y_{\rm nuc}^{(\mu)}(F)}\,\frac{\De E_{F(\mu)}^{(3)}}{\mub B\mf}\,,
\end{equation}
where
\begin{equation}
\begin{split}
\label{DeECC} 
\De E_{F(\mu)}^{(3)}=
\sum_{\mj\mi}\sum_{\mj'\mi'}C^{F\mf}_{\frac{1}{2}\mj I\mi}
C^{F\mf}_{\frac{1}{2}\mj' I\mi'}
\la I\mi|I_{\mu}^{(3a)}+I_{\mu}^{(3b)}
+I_{\mu}^{(3c)}+I_{\mu}^{(3d)} 
|I\mi'\ra \,,
\end{split}
\end{equation}
\begin{align}
\label{Ia}
I_{\mu}^{(3a)}&= \sum_{\eps_c=\eps_{1s}} 
\biggl(
\sum_{n_1,n_2}^{(\eps_{n_1}\neq \eps_v,\eps_{n_2}\neq \eps_v)}
\frac{2}{(\eps_v-\eps_{n_1})(\eps_v-\eps_{n_2})}
\biggl[
\la v|V_{\vecb}^{(e)}|n_1\ra\la n_1|V_{\rm HFS}^{(\mu)}|n_2\ra
\la n_2 c|I(0)|v' c\ra  
 \nonumber \\
&+\la v|V_{\vecb}^{(e)}|n_1\ra\la n_1 c|I(0)|n_2 c\ra
\la n_2|V_{\rm HFS}^{(\mu)}|v'\ra
 +\la v|V_{\rm HFS}^{(\mu)}|n_1\ra\la n_1|V_{\vecb}^{(e)}|n_2\ra
\la n_2 c|I(0)|v' c\ra
\biggr]
\nonumber \\
&-\sum_{\eps_{\tv}=\eps_v}\sum_{n}^{(\eps_{n}\neq \eps_v)}
\frac{2}{(\eps_v-\eps_{n})^2}
\biggl[
\la v|V_{\vecb}^{(e)}|n\ra\la n|V_{\rm HFS}^{(\mu)}|\tv\ra
\la \tv c|I(0)|v' c\ra
\nonumber \\
&+\la v|V_{\vecb}^{(e)}|n\ra\la n c|I(0)|\tv c\ra
\la \tv|V_{\rm HFS}^{(\mu)}|v'\ra
+\la v|V_{\vecb}^{(e)}|\tv\ra\la \tv|V_{\rm HFS}^{(\mu)}|n\ra
\la n c|I(0)|v' c\ra
\biggr]
 \biggr)\,,
\\
\label{Ib}
I_{\mu}^{(3b)}&= -\sum_{\eps_c=\eps_{1s}} 
\biggl(
\sum_{n_1,n_2}^{(\eps_{n_1}\neq \eps_v,\eps_{n_2}\neq \eps_v)}
\frac{2}{(\eps_v-\eps_{n_1})(\eps_v-\eps_{n_2})}
\biggl[
\la v|V_{\vecb}^{(e)}|n_1\ra\la n_1|V_{\rm HFS}^{(\mu)}|n_2\ra
\la n_2 c|I(\om)|c v'\ra  
 \nonumber \\
&+\la v|V_{\vecb}^{(e)}|n_1\ra\la n_1 c|I(\om)|c n_2\ra
\la n_2|V_{\rm HFS}^{(\mu)}|v'\ra
 +\la v|V_{\rm HFS}^{(\mu)}|n_1\ra\la n_1|V_{\vecb}^{(e)}|n_2\ra
\la n_2 c|I(\om)|c v'\ra
\biggr]
\nonumber \\
&-\sum_{\eps_{\tv}=\eps_v}\sum_{n}^{(\eps_{n}\neq \eps_v)}
\frac{2}{(\eps_v-\eps_{n})^2}
\biggl[
\la v|V_{\vecb}^{(e)}|n\ra\la n|V_{\rm HFS}^{(\mu)}|\tv\ra
\la \tv c|I(\om)|c v'\ra
\nonumber \\
&+\la v|V_{\vecb}^{(e)}|n\ra\la n c|I(\om)|c \tv\ra
\la \tv|V_{\rm HFS}^{(\mu)}|v'\ra
+\la v|V_{\vecb}^{(e)}|\tv\ra\la \tv|V_{\rm HFS}^{(\mu)}|n\ra
\la n c|I(\om)|c v'\ra
\biggr]
\nonumber \\
&+\sum_{\eps_{\tv}=\eps_v}\sum_{n}^{(\eps_{n}\neq \eps_v)}
\frac{2}{\eps_v-\eps_{n}}
\biggl[
\la v|V_{\vecb}^{(e)}|n\ra\la n|V_{\rm HFS}^{(\mu)}|\tv\ra
\la \tv c|I'(\om)|c v'\ra
\nonumber \\
&+\la v|V_{\vecb}^{(e)}|n\ra\la n c|I'(\om)|c \tv\ra
\la \tv|V_{\rm HFS}^{(\mu)}|v'\ra
+\la v|V_{\vecb}^{(e)}|\tv\ra\la \tv|V_{\rm HFS}^{(\mu)}|n\ra
\la n c|I'(\om)|c v'\ra
\biggr]
\nonumber \\
&+\sum_{\eps_{\tv}=\eps_v}\sum_{\eps_{\bv}=\eps_v}
\la v|V_{\vecb}^{(e)}|\tv\ra\la \tv c|I''(\om)|c \bv\ra
\la \bv|V_{\rm HFS}^{(\mu)}|v'\ra 
\biggr)\,,
\\
\label{Ic}
I_{\mu}^{(3c)}&= \sum_{\eps_c=\eps_{1s}} 
\biggl(
\sum_{n_1,n_2}^{(\eps_{n_1}\neq \eps_v,\eps_{n_2}\neq \eps_c)}
\frac{2}{(\eps_v-\eps_{n_1})(\eps_c-\eps_{n_2})}
\biggl[
\la v|V_{\vecb}^{(e)}|n_1\ra\la c|V_{\rm HFS}^{(\mu)}|n_2\ra
\la n_1 n_2|I(0)|v' c\ra  
\nonumber \\
&+\la v|V_{\vecb}^{(e)}|n_1\ra\la n_1 c|I(0)|v' n_2\ra
\la n_2|V_{\rm HFS}^{(\mu)}|c\ra
+\la v|V_{\rm HFS}^{(\mu)}|n_1\ra\la c|V_{\vecb}^{(e)}|n_2\ra
\la n_1 n_2|I(0)|v' c\ra
\nonumber \\  
&+\la v|V_{\rm HFS}^{(\mu)}|n_1\ra\la n_1 c|I(0)|v' n_2\ra
\la n_2|V_{\vecb}^{(e)}|c\ra
\biggr]
\nonumber \\
&+\sum_{n_1,n_2}^{(\eps_{n_1}\neq \eps_c,\eps_{n_2}\neq \eps_c)}
\frac{2}{(\eps_c-\eps_{n_1})(\eps_c-\eps_{n_2})}
\biggl[
\la c|V_{\vecb}^{(e)}|n_1\ra
\la n_1|V_{\rm HFS}^{(\mu)}|n_2\ra\la n_2 v|I(0)|c v'\ra
\nonumber \\
&+\la c|V_{\vecb}^{(e)}|n_1\ra\la n_1 v|I(0)|n_2 v'\ra 
\la n_2|V_{\rm HFS}^{(\mu)}|c\ra 
+\la c|V_{\rm HFS}^{(\mu)}|n_1\ra\la n_1|V_{\vecb}^{(e)}|n_2\ra
\la n_2 v|I(0)|c v'\ra
\biggr]
\nonumber \\
&-\sum_{\eps_{\tc}=\eps_{1s}}\sum_{n}^{(\eps_{n}\neq \eps_c)}
\frac{2}{(\eps_c-\eps_{n})^2}
\biggl[
\la c|V_{\vecb}^{(e)}|n\ra\la n|V_{\rm HFS}^{(\mu)}|\tc\ra
\la \tc v|I(0)|c v'\ra
\nonumber \\
&+\la c|V_{\vecb}^{(e)}|n\ra\la n v|I(0)|\tc v'\ra
\la \tc|V_{\rm HFS}^{(\mu)}|c \ra
+\la c|V_{\vecb}^{(e)}|\tc\ra\la \tc|V_{\rm HFS}^{(\mu)}|n \ra
\la n v|I(0)|c v'\ra
\biggr]
 \biggr)\,,
\end{align}
\begin{align}\label{Id}
I_{\mu}^{(3d)}&= -\sum_{\eps_c=\eps_{1s}} 
\biggl(
\sum_{n_1,n_2}^{(\eps_{n_1}\neq \eps_v,\eps_{n_2}\neq \eps_c)}
\frac{2}{(\eps_v-\eps_{n_1})(\eps_c-\eps_{n_2})}
\biggl[
\la v|V_{\vecb}^{(e)}|n_1\ra\la c|V_{\rm HFS}^{(\mu)}|n_2\ra
\la n_1 n_2|I(\om)|c v'\ra 
\nonumber \\
&+\la v|V_{\vecb}^{(e)}|n_1\ra\la n_1 c|I(\om)|n_2 v'\ra 
\la n_2|V_{\rm HFS}^{(\mu)}|c\ra
+\la v|V_{\rm HFS}^{(\mu)}|n_1\ra\la c|V_{\vecb}^{(e)}|n_2\ra
\la n_1 n_2|I(\om)|c v'\ra 
\nonumber \\
&+\la v|V_{\rm HFS}^{(\mu)}|n_1\ra\la n_1 c|I(\om)|n_2 v'\ra 
\la n_2|V_{\vecb}^{(e)}|c\ra
\biggr]
\nonumber \\
&+\sum_{n_1,n_2}^{(\eps_{n_1}\neq \eps_c,\eps_{n_2}\neq \eps_c)}
\frac{2}{(\eps_c-\eps_{n_1})(\eps_c-\eps_{n_2})}
\biggl[
\la c|V_{\vecb}^{(e)}|n_1\ra\la n_1|V_{\rm HFS}^{(\mu)}|n_2\ra
\la n_2 v|I(\om)|v' c\ra 
\nonumber \\
&+\la c|V_{\vecb}^{(e)}|n_1\ra\la n_1 v|I(\om)|v' n_2\ra
\la n_2|V_{\rm HFS}^{(\mu)}|c\ra
+\la c|V_{\rm HFS}^{(\mu)}|n_1\ra\la n_1|V_{\vecb}^{(e)}|n_2\ra
\la n_2 v|I(\om)|v' c\ra 
\biggr]
\nonumber \\
&-\sum_{\eps_{\tc}=\eps_{1s}}\sum_{n}^{(\eps_{n}\neq \eps_c)}
\frac{2}{(\eps_c-\eps_{n})^2}
\biggl[
\la c|V_{\vecb}^{(e)}|n \ra\la n|V_{\rm HFS}^{(\mu)}|\tc\ra
\la \tc v|I(\om)|v' c\ra 
\nonumber \\
&+\la c|V_{\vecb}^{(e)}|n \ra
\la n v|I(\om)|v' \tc\ra \la \tc|V_{\rm HFS}^{(\mu)}|c\ra
+\la c|V_{\vecb}^{(e)}|\tc \ra\la \tc|V_{\rm HFS}^{(\mu)}|n\ra
\la n v|I(\om)|v' c\ra
\biggr]
\nonumber \\
&-\sum_{\eps_{\tc}=\eps_{1s}}\sum_{n}^{(\eps_{n}\neq \eps_c)}
\frac{2}{\eps_c-\eps_{n}}
\biggl[
\la c|V_{\vecb}^{(e)}|n\ra\la n|V_{\rm HFS}^{(\mu)}|\tc\ra
\la \tc v|I'(\om)|v' c\ra
\nonumber \\
&+\la c|V_{\vecb}^{(e)}|n\ra
\la n v|I'(\om)|v' \tc\ra\la \tc|V_{\rm HFS}^{(\mu)}|c\ra
+\la c|V_{\vecb}^{(e)}|\tc\ra\la \tc|V_{\rm HFS}^{(\mu)}|n\ra
\la n v|I'(\om)|v' c\ra
\biggr]
\nonumber \\
&-\sum_{\eps_{\tc}=\eps_{1s}}\sum_{n}^{(\eps_{n}\neq \eps_v)}
\frac{2}{\eps_v-\eps_{n}}
\biggl[
\la v|V_{\vecb}^{(e)}|n\ra\la n c|I'(\om)|\tc v'\ra
\la \tc|V_{\rm HFS}^{(\mu)}|c\ra
+\la v|V_{\rm HFS}^{(\mu)}|n\ra\la n c|I'(\om)|\tc v'\ra
\la \tc|V_{\vecb}^{(e)}|c\ra
\biggr]
\nonumber \\
&+\sum_{\eps_{\tv}=\eps_v}\sum_{n}^{(\eps_{n}\neq \eps_c)}
\frac{2}{\eps_c-\eps_{n}}
\biggl[
\la v|V_{\vecb}^{(e)}|\tv\ra\la \tv c|I'(\om)|n v'\ra
\la n|V_{\rm HFS}^{(\mu)}|c\ra
+\la v|V_{\rm HFS}^{(\mu)}|\tv\ra\la \tv c|I'(\om)|n v'\ra
\la n|V_{\vecb}^{(e)}|c\ra
\biggr]
\nonumber \\
&-\sum_{\eps_{\tv}=\eps_v}\sum_{\eps_{\tc}=\eps_{1s}}
\biggl[
\la v|V_{\vecb}^{(e)}|\tv\ra\la \tv c|I''(\om)|\tc v'\ra
\la \tc|V_{\rm HFS}^{(\mu)}|c\ra
+\la v|V_{\rm HFS}^{(\mu)}|\tv\ra\la \tv c|I''(\om)|\tc v'\ra
\la \tc|V_{\vecb}^{(e)}|c\ra
\biggr]
\nonumber \\
&+\sum_{\eps_{\tc}=\eps_{1s}}\sum_{\eps_{\bc}=\eps_{1s}}
\la c|V_{\vecb}^{(e)}|\tc\ra\la \tc v|I''(\om)|v' \bc\ra
\la \bc|V_{\rm HFS}^{(\mu)}|c\ra
 \biggr)\,.
\end{align}
Here
\begin{equation}
\la n_1 n_2|I(\om)|n_3 n_4 \ra \equiv \int
d{\vec{x}}_1 d{\vec{x}}_2 u_{n_1}^{\dag}({\vec{x}}_1)
u_{n_2}^{\dag}({\vec{x}}_2)
I(\om) u_{n_3}({\vec{x}}_1)u_{n_4}({\vec{x}}_2)\,,
\end{equation}
\begin{equation}
I(\om)=\alpha\frac{(1-\vec{\alpha}_1\cdot\vec{\alpha}_2)
\cos(\omega r_{12})}{r_{12}}\,,
\end{equation}
\begin{equation}
I'(\om)=\frac{dI(\om)}{d\om}\,,
\,I''(\om)=\frac{d^2 I(\om)}{d{\om}^2}\,,
\end{equation}
$\omega=\eps_v-\eps_{1s}\,$, and $r_{12}=|{\vec{x}}_1-{\vec{x}}_2|$.

For checking purposes the corresponding calculation of the 
function $B_{\mu}(\az)$ was performed in the Coulomb gauge as well.
The results of both calculations coincide with each other.

\section{\large Numerical results}

In Table 1, we present the numerical results for the function 
$S_2(\alpha Z)$. The exact values calculated for the point-like and 
extended nuclear charge distribution
models are presented in the fourth and fifth columns, respectively.

In Table 2, we present the numerical results for the function
$B_{\mu}(\az)$ defined by Eq. (\ref{defBmu}),
\begin{equation}
B_{\mu}(\az)= B_{\mu}^{(a)}(\az)+B_{\mu}^{(b)}(\az)
+B_{\mu}^{(c)}(\az)+B_{\mu}^{(d)}(\az)\,,
\end{equation}
for the $2s$ state. $B_{\mu}^{(a)}(\az)$, $B_{\mu}^{(b)}(\az)$,
$B_{\mu}^{(c)}(\az)$, and $B_{\mu}^{(d)}(\az)$ denote contributions
from the corresponding diagrams presented in Fig. \ref{Diagrams3}.
All the values are calculated for the extended nuclear charge
distribution. The root-mean-square nuclear charge radii were taken from
\cite{RnewAng}. For those elements for which no accurate
experimental radii were available we employed the empirical
expression \cite{joh85}
\begin{align}
\langle r^2\rangle^{1/2}=0.836 A^{1/3}+0.570(\pm 0.05)\
\text{fm}\,,
\end{align}
where $A$ is the nuclear mass expressed in a.m.u.  

The dual
kinetic balance (DKB) approach to basis-set expansion for the
Dirac equation \cite{DKB04} was used for these calculations. 
The basis DKB functions were constructed from B-splines \cite{joh86,
joh88}. The Fermi model was used for the nuclear charge distribution. 
The uncertainties were estimated by adding quadratically two errors,
one obtained by varying $\la r^2\ra^{1/2}$ within its uncertainty 
and the other obtained by changing the model of the nuclear-charge 
distribution from the Fermi to the homogeneously-charged-sphere 
model.

\section{\large Discussion}

The total $2s$ $g$-factor value of a lithiumlike ion with nonzero
nuclear spin can be represented by
\begin{eqnarray}
g&=&(g_{\rm D}+\Delta g_{\rm int}+
\Delta g_{\rm QED}+\Delta g_{\rm rec}^{(e)}+ \Delta
g_{\rm NS}+\Delta g_{\rm NP}) Y_{\rm el}(F)
 \nonumber \\
&&-\frac{\me}{\mpr}(g_I+\Delta g_{\rm rec}^{(n)}) Y_{\rm
nuc}^{(\mu)}(F)+\delta g_{\rm HFS(\mu)}^{(2s)} 
+\delta g_{\rm HFS(Q)}^{(2s)}\,,
\end{eqnarray}
where
\begin{eqnarray} \label{Yel}
Y_{\rm el}(F)=\frac{F(F+1)+3/4-I(I+1)}{2F(F+1)}=
\begin{cases}
  -\frac{1}{2I+1}   &\text{for $F=I-\frac{1}{2}$}\\
  \frac{1}{2I+1}        &\text{for $F=I+\frac{1}{2}$}
\end{cases}\,,
\end{eqnarray}
$Y_{\rm nuc}^{(\mu)}(F)$ is defined by equation (\ref{Ynucdip}).
The individual contributions to the $g$ factor of the ground state 
of some Li-like ions are presented in Table 3 for $F=I-1/2$ 
and in Table 4 for $F=I+1/2$. 
The Dirac point-nucleus value is obtained by
Eq. (\ref{gD}). The interelectronic-interaction ($\De g_{\rm int}$), 
QED ($\Delta g_{\rm QED}$), nuclear-recoil ($\Delta g_{\rm rec}^{(e)}$), 
and nuclear-size ($\Delta g_{\rm NS}$) corrections are obtained as described
in Refs. \cite{glasha04,glavol06}. To
estimate the nuclear-polarization correction to the $g$ factor of
Li-like ions with nonzero nuclear spin, we used the corresponding values
for the zero-nuclear-spin isotopes \cite{nef02} with the 100\%
uncertainty. This correction is essential only for heavy elements.
Since in all the cases under consideration the absolute value of the recoil
correction $\Delta g_{\rm rec}^{(n)}$ to the bound-nucleus 
$g$ factor is smaller than
$10^{-11}$ \cite{mar01}, it is omitted in Tables 3 and 4.

The hyperfine-interaction corrections
$\delta g_{\rm HFS(\mu)}^{(2s)}$ and $\delta g_{\rm HFS(Q)}^{(2s)}$ 
are given by formulas (\ref{gcorrdip}) and (\ref{gcorrquadr}), respectively.
The uncertainty due to uncalculated second- and higher-order terms 
in Eq. (\ref{gcorrdip}) was estimated as the first-order correction 
($\sim B_{\mu}(\az)/Z$) multiplied by the factor $2/Z$. The
uncertainty due to uncalculated first- and higher-order terms 
in Eq. (\ref{gcorrquadr}) was estimated in a similar way.

It can be seen from Tables 3 and 4 that, as a rule, the
electric-quadrupole hyperfine-interaction correction is much
smaller than the magnetic-dipole one.
This is due to an additional factor $(\alpha Z)^2$ in formula
(\ref{gcorrquadr}) compared to formula (\ref{gcorrdip}) and small
values of $Q$ for low-$Z$ ions. However, in case of $^{235}{\rm
U}^{89+}$ these corrections are of the same order of magnitude.

The uncertainties of the nuclear magnetic moments indicated in
Tables 3 and 4, as a rule, do not include errors due to unknown chemical
shifts which, in some cases, can contribute on the level of a few
tenths percents. This means that measurements of the $g$ factor
with the aforementioned accuracy could provide the most accurate
determinations of the nuclear magnetic moments.
The hyperfine-interaction correction  evaluated
in this paper will be important for this determination.

\section*{\large Acknowledgements}

D.L.M. thanks N.S. Oreshkina for valuable advice.
He is also grateful to GSI.
This work was supported by INTAS-GSI (Grant No. 05-111-4937).
V.M.S. acknowledges the support by INTAS-GSI grant No. 06-1000012-8881.

\newpage
\begin{table}
\caption{Numerical results for the function $S_2(\az)$
defined by Eq. (\ref{defS2}). 
$S^{\rm point}_2$ is the point-nucleus value
obtained by formula (\ref{S2}). $S^{\rm ext}_2$ is the 
extended-nucleus value. The values of $\la r^2\ra^{1/2}$ are taken
from Ref. \cite{RnewAng}. }
\begin{center}
\begin{tabular}{||l|l|l|l|l||}
\hline Ion &  $Z$&  $\la r^2\ra^{1/2}$ (fm)&    
$S^{\rm point}_2(\az)$ &   $S^{\rm ext}_2(\az)$
\\ \hline
$^{7}\rm{Li}$          &  3 & 2.431 &1.00076 &1.00076   
\\ \hline
$^{9}\rm{Be}^{+}$      &  4 & 2.518 &1.00136 &1.00136    
\\ \hline
$^{11}\rm{B}^{2+}$     &  5 & 2.406 &1.00212 &1.00212   
\\ \hline
$^{13}\rm{C}^{3+}$     &  6 & 2.461 &1.00306 &1.00306   
\\ \hline
$^{14}\rm{N}^{4+}$     &  7 & 2.558 &1.00417 &1.00417   
\\ \hline
$^{17}\rm{O}^{5+}$     &  8 & 2.695 &1.00545 &1.00545   
\\ \hline
$^{19}\rm{F}^{6+}$     &  9 & 2.898 &1.00691 &1.00691    
\\ \hline
$^{21}\rm{Ne}^{7+}$    & 10 & 2.967 &1.00855 &1.00854    
\\ \hline
$^{33}\rm{S}^{13+}$    & 16 & 3.251 &1.02224 &1.02218   
\\ \hline
$^{43}\rm{Ca}^{17+}$   & 20 & 3.493 &1.03527 &1.03513   
\\ \hline
$^{53}\rm{Cr}^{21+}$   & 24 & 3.659 &1.05171 &1.05145    
\\ \hline
$^{67}\rm{Zn}^{27+}$   & 30 & 3.964 &1.08359 &1.08296(1)
\\ \hline
$^{73}\rm{Ge}^{29+}$   & 32 & 4.063 &1.09637 &1.09555(2) 
\\ \hline
$^{91}\rm{Zr}^{37+}$   & 40 & 4.284 &1.16037 &1.15824(3)
\\ \hline
$^{113}\rm{In}^{46+}$  & 49 & 4.602 &1.26402 &1.25820(6) 
\\ \hline
$^{115}\rm{In}^{46+}$  & 49 & 4.617 &1.26402 &1.25818(6)
\\ \hline
$^{119}\rm{Sn}^{47+}$  & 50 & 4.645 &1.27819 &1.27170(7) 
\\ \hline
$^{127}\rm{I}^{50+}$   & 53 & 4.750 &1.32460 &1.31566(9) 
\\ \hline
$^{129}\rm{Xe}^{51+}$  & 54 & 4.776 &1.34148 &1.33156(10)
\\ \hline
$^{131}\rm{Xe}^{51+}$  & 54 & 4.781 &1.34148 &1.33155(10)
\\ \hline
$^{143}\rm{Nd}^{57+}$  & 60 & 4.923 &1.46042 &1.4420(2)  
\\ \hline
$^{159}\rm{Tb}^{62+}$  & 65 & 5.060 &1.58862 &1.5577(8)  
\\ \hline
$^{173}\rm{Yb}^{67+}$  & 70 & 5.304 &1.75329 &1.7004(3)  
\\ \hline
$^{177}\rm{Hf}^{69+}$  & 72 & 5.333 &1.83238 &1.7674(4)  
\\ \hline
$^{185}\rm{Re}^{72+}$  & 75 & 5.329 &1.96892 &1.8806(6)  
\\ \hline
$^{187}\rm{Re}^{72+}$  & 75 & 5.339 &1.96892 &1.8805(6)  
\\ \hline
$^{195}\rm{Pt}^{75+}$  & 78 & 5.428 &2.13172 &2.0095(7)  
\\ \hline
$^{197}\rm{Au}^{76+}$  & 79 & 5.436 &2.19299 &2.0570(7) 
\\ \hline
$^{199}\rm{Hg}^{77+}$  & 80 & 5.448 &2.25827 &2.1068(8) 
\\ \hline
$^{207}\rm{Pb}^{79+}$  & 82 & 5.494 &2.40235 &2.2136(10) 
\\ \hline
$^{209}\rm{Bi}^{80+}$  & 83 & 5.521 &2.48199 &2.2709(10) 
\\ \hline
$^{229}\rm{Th}^{87+}$  & 90 & 5.681 &3.23718 &2.764(3)
\\ \hline
$^{231}\rm{Pa}^{88+}$  & 91 & 5.700 &3.38354 &2.850(3)   
\\ \hline
$^{235}\rm{U}^{89+}$   & 92 & 5.829 &3.54308 &2.935(2)   
\\ \hline
$^{257}\rm{Fm}^{97+}$  &100 & 5.886 &5.57470 &3.870(7)   
\\ \hline
\end{tabular}
\end{center}
\end{table}
\begin{table}
\caption{The contributions to the interelectronic-interaction correction 
$B_{\mu}(\az)$, defined by Eq. (\ref{defBmu}), from the diagrams presented
in Figs. 3--5. The values of $\la r^2\ra^{1/2}$ are given
in Table 1.}
\begin{center}
\begin{tabular}{||l|l|l|l|l|l|l||}
\hline Ion &  $Z$&  $B^{(a)}_{\mu}(\az)$  &  $ B^{(b)}_{\mu}(\az)$  &   $ B^{(c)}_{\mu}(\az)$ &   
 $ B^{(d)}_{\mu}(\az)$ &    $ B_{\mu}(\az)$
\\ \hline
$^{7}\rm{Li}$          &  3   &-1.45058   &-0.0588955   &-0.230825   &0.146787   &-1.59351
\\ \hline
$^{9}\rm{Be}^{+}$      &  4   &-1.45215   &-0.0591841   &-0.231117   &0.147153   &-1.59529
\\ \hline
$^{11}\rm{B}^{2+}$     &  5   &-1.45417   &-0.0595557   &-0.231493   &0.147625   &-1.59759  
\\ \hline
$^{13}\rm{C}^{3+}$     &  6   &-1.45664   &-0.0600107   &-0.231953   &0.148205   &-1.60040
\\ \hline
$^{14}\rm{N}^{4+}$     &  7   &-1.45957   &-0.0605498   &-0.232499   &0.148894   &-1.60373
\\ \hline
$^{17}\rm{O}^{5+}$     &  8   &-1.46296   &-0.0611736   &-0.233130   &0.149694   &-1.60757
\\ \hline
$^{19}\rm{F}^{6+}$     &  9   &-1.46682   &-0.0618828   &-0.233850   &0.150607   &-1.61194
\\ \hline
$^{21}\rm{Ne}^{7+}$    & 10   &-1.47114   &-0.0626786   &-0.234657   &0.151636   &-1.61684
\\ \hline
$^{33}\rm{S}^{13+}$    & 16   &-1.50734(1)&-0.0693364(4)&-0.241442(1)&0.160434(1)&-1.65769(1)
\\ \hline
$^{43}\rm{Ca}^{17+}$   & 20   &-1.54187(1)&-0.0756850(2)&-0.247960(1)&0.169119(2)&-1.69639(1)
\\ \hline
$^{53}\rm{Cr}^{21+}$   & 24   &-1.58556(2)&-0.0837272(8)&-0.256270(2)&0.180504(4)&-1.74505(2)
\\ \hline
$^{67}\rm{Zn}^{27+}$   & 30   &-1.67053(4)&-0.099395(3) &-0.272617(4)&0.203796(9)&-1.83874(3)
\\ \hline
$^{73}\rm{Ge}^{29+}$   & 32   &-1.70468(4)&-0.105707(4) &-0.279253(6)&0.21355(1) &-1.87609(4)
\\ \hline
$^{91}\rm{Zr}^{37+}$   & 40   &-1.87649(9)&-0.13762(1)  &-0.31312(1) &0.26566(3) &-2.06157(9)
\\ \hline
$^{113}\rm{In}^{46+}$  & 49   &-2.1556(2) &-0.19021(3)  &-0.36966(3) &0.35919(5) &-2.3563(2) 
\\ \hline
$^{115}\rm{In}^{46+}$  & 49   &-2.1556(2) &-0.19021(3)  &-0.36965(3) &0.35917(5) &-2.3563(2)      
\\ \hline
$^{119}\rm{Sn}^{47+}$  & 50   &-2.1938(2) &-0.19748(3)  &-0.37751(3) &0.37268(7) &-2.3961(2)
\\ \hline
$^{127}\rm{I}^{50+}$   & 53   &-2.3186(2) &-0.22141(4)  &-0.40340(5) &0.41789(9)&-2.5255(3)
\\ \hline
$^{129}\rm{Xe}^{51+}$  & 54   &-2.3640(3) &-0.23017(4)  &-0.41287(5) &0.43469(9)&-2.5724(3)
\\ \hline
$^{131}\rm{Xe}^{51+}$  & 54   &-2.3640(3) &-0.23016(4)  &-0.41287(5) &0.43468(9)&-2.5723(3)
\\ \hline
$^{143}\rm{Nd}^{57+}$  & 60   &-2.6825(5) &-0.29243(8)  &-0.48024(8) &0.5574(2)  &-2.8977(3)
\\ \hline
$^{159}\rm{Tb}^{62+}$  & 65   &-3.021(2) &-0.3603(3) &-0.5535(4)&0.6961(8)  &-3.239(2)
\\ \hline
$^{173}\rm{Yb}^{67+}$  & 70   &-3.4456(9) &-0.4477(2)   &-0.6470(2)  &0.8792(4)  &-3.6612(10)
\\ \hline
$^{177}\rm{Hf}^{69+}$  & 72   &-3.647(2)  &-0.4900(2)   &-0.6919(2)  &0.9690(5)  &-3.8597(12)
\\ \hline
$^{185}\rm{Re}^{72+}$  & 75   &-3.990(2)  &-0.5632(3)   &-0.7692(4)  &1.1257(6)  &-4.196(2)
\\ \hline
$^{187}\rm{Re}^{72+}$  & 75   &-3.989(2)  &-0.5632(3)   &-0.7691(4)  &1.1256(6)  &-4.196(2)
\\ \hline
$^{195}\rm{Pt}^{75+}$  & 78   &-4.383(2)  &-0.6494(4)   &-0.8591(4)  &1.3116(8) &-4.580(2)
\\ \hline
$^{197}\rm{Au}^{76+}$  & 79   &-4.529(2)  &-0.6818(4)   &-0.8927(4)  &1.3819(9) &-4.722(3)
\\ \hline
$^{199}\rm{Hg}^{77+}$  & 80   &-4.683(3)  &-0.7162(5)   &-0.9282(5)  &1.4565(10) &-4.870(3)
\\ \hline
$^{207}\rm{Pb}^{79+}$  & 82   &-5.013(3)  &-0.7912(5)   &-1.0052(7)  &1.6196(12) &-5.189(3)
\\ \hline
$^{209}\rm{Bi}^{80+}$  & 83   &-5.191(4)  &-0.8322(5)   &-1.0469(7)  &1.7089(13) &-5.361(3)
\\ \hline
$^{229}\rm{Th}^{87+}$  & 90   &-6.738(9)  &-1.2047(16)  &-1.417(2) &2.520(4)   &-6.840(9)
\\ \hline
$^{231}\rm{Pa}^{88+}$  & 91   &-7.011(11)  &-1.2732(17)  &-1.484(2) &2.669(5)   &-7.099(9)
\\ \hline
$^{235}\rm{U}^{89+}$   & 92   &-7.279(7)  &-1.3429(13)  &-1.5499(15) &2.819(4)   &-7.353(7)
\\ \hline
$^{257}\rm{Fm}^{97+}$  &100   &-10.26(3)&-2.156(5)    &-2.301(5)   &4.558(12)   &-10.16(2)
\\ \hline
\end{tabular}
\end{center}
\end{table}
\begin{table}
\caption{The individual contributions to the ground-state $g$ factor of
lithiumlike ions with nonzero nuclear spin for $F=I-\frac{1}{2}$.
The values of $\la r^2\ra^{1/2}$ are given in Table 1. 
The values of $I$, $\frac{\mu}{\mun}$, and $Q$ are
given in Table 4. }
\begin{center}
\begin{tabular}{||c|r@{.}l|r@{.}l|r@{.}l|r@{.}l||}
\hline
 Ion             
& \multicolumn{2}{c|}{ $^{17}\rm{O}^{5+}$}
&\multicolumn{2}{c|}{$^{33}\rm{S}^{13+}$}
&\multicolumn{2}{c|}{$^{43}\rm{Ca}^{17+}$}
&\multicolumn{2}{c||}{$^{53}\rm{Cr}^{21+}$}
\\ \hline

$g_{\rm D}Y_{\rm{el}}(F)$
&-0&333238563      &-0&499429548
&-0&249553251      &-0&498709516
\\ \hline

$\De g_{\rm{int}}Y_{\rm{el}}(F)$
&-0&000029443(5)      &-0&00009031(2)
&-0&000056806(18)     &-0&00013712(5)
\\ \hline

$\De g_{\rm{QED}}Y_{\rm{el}}(F)$
&-0&000386592(2)      &-0&000580176(15)
&-0&000290213(13)     &-0&00058076(4)
\\ \hline

$\De g_{\rm{rec}}^{(e)}Y_{\rm{el}}(F)$
&-0&000000003     &-0&000000011(1)
&-0&000000007     &-0&000000017(1)
\\ \hline

$\De g_{\rm{NS}}Y_{\rm{el}}(F)$
&0&0            &-0&000000001
&-0&000000002   &-0&000000009
\\ \hline

$-\frac{\me}{\mpr}g_{I}Y_{\rm{nuc}}^{(\mu)}(F)$
&0&00048132(2)      &-0&000292197(1)
&0&000230660(1)     & 0&00021537(1)
\\ \hline

$\delta g_{\rm{HFS}(\mu)}^{(2s)}$
&-0&000000014(1)     & 0&000000019
&-0&000000019        &-0&000000022
\\ \hline

$\delta g_{\rm{HFS}(Q)}^{(2s)}$
&0&0     &0&0
&0&0     &0&0
\\ \hline

$g$
&-0&33317330(2)    &-0&50039222(3)
&-0&24966964(2)    &-0&49921208(7)
\\ \hline
\end{tabular}
\end{center}
\begin{center}
\begin{tabular}{||c|r@{.}l|r@{.}l|r@{.}l|r@{.}l||}
\hline
 Ion 
&\multicolumn{2}{c|}{$^{73}\rm{Ge}^{29+}$}
&\multicolumn{2}{c|}{$^{131}\rm{Xe}^{51+}$}
&\multicolumn{2}{c|}{$^{209}\rm{Bi}^{80+}$}
&\multicolumn{2}{c||}{$^{235}\rm{U}^{89+}$} 
\\ \hline

$g_{\rm D}Y_{\rm{el}}(F)$ 
&-0&199075231   &-0&493187551 
&-0&193006882   &-0&238840328 
\\ \hline

$\De g_{\rm{int}}Y_{\rm{el}}(F)$
&-0&00007397(4)       &-0&0003249(3)
&-0&0002175(3)      &-0&0003127(5)
\\ \hline

$\De g_{\rm{QED}}Y_{\rm{el}}(F)$
&-0&00023270(2)   &-0&00058679(11) 
&-0&00024077(13)  &-0&0003049(2)
\\ \hline

$\De g_{\rm{rec}}^{(e)}Y_{\rm{el}}(F)$
&-0&000000009(1)  &-0&000000040(15)
&-0&00000003(4) &-0&00000004(9) 
\\ \hline

$\De g_{\rm{NS}}Y_{\rm{el}}(F)$
&-0&000000016      &-0&000000839(3)
&-0&000008752(17)  &-0&00002999(6)
\\ \hline

$\De g_{\rm{NP}}Y_{\rm{el}}(F)$
&\multicolumn{2}{c|}{------}
&\multicolumn{2}{c|}{------}
&\multicolumn{2}{c|}{------}
&0&00000003(3) 
\\ \hline

$-\frac{\me}{\mpr}g_{I}Y_{\rm{nuc}}^{(\mu)}(F)$
& 0&000117082   &-0&000314000(2)
&-0&00054724(3) & 0&000068(12)
\\ \hline

$\delta g_{\rm{HFS}(\mu)}^{(2s)}$
&-0&000000017   & 0&000000097 
& 0&000000445   &-0&000000080(14)
\\ \hline

$\delta g_{\rm{HFS}(Q)}^{(2s)}$
&0&0         & 0&000000001 
&0&000000002 &-0&000000046(1) 
\\ \hline

$g$
&-0&19926487(4)  &-0&4944140(3) 
&-0&1940207(3) &-0&239420(12) 
\\ \hline
\end{tabular}
\end{center}
\end{table}

\begin{table}
\caption{ The individual contributions to the ground-state $g$ factor of
lithiumlike ions with nonzero nuclear spin for $F=I+\frac{1}{2}$.
The values of $\la r^2\ra^{1/2}$ are given in Table 1. 
The values of $\frac{\mu}{\mun}$ and $Q$ are
taken from Refs. \cite{rag89} and \cite{PPyy01}, respectively.}
\begin{center}
\begin{tabular}{||c|r@{.}l|r@{.}l|r@{.}l|r@{.}l||}
\hline
 Ion
&\multicolumn{2}{c|}{$^{13}\rm{C}^{3+}$}
&\multicolumn{2}{c|}{$^{17}\rm{O}^{5+}$}
&\multicolumn{2}{c|}{$^{33}\rm{S}^{13+}$}
&\multicolumn{2}{c||}{$^{43}\rm{Ca}^{17+}$} 
\\ \hline

$I$
&\multicolumn{2}{c|}{1/2} 
&\multicolumn{2}{c|}{5/2}
&\multicolumn{2}{c|}{3/2}
&\multicolumn{2}{c||}{7/2} 
\\ \hline

$\mu/\mun$
&\multicolumn{2}{c|}{0.7024118(14)}
&\multicolumn{2}{c|}{-1.89379(9)}
&\multicolumn{2}{c|}{0.6438212(14)}
&\multicolumn{2}{c||}{-1.317643(7)}
\\ \hline

$Q$ (barn)
&\multicolumn{2}{c|}{------}
&\multicolumn{2}{c|}{-0.02558(22)}
&\multicolumn{2}{c|}{-0.0678(13)}
&\multicolumn{2}{c||}{-0.0408(8)}
\\ \hline

$g_{\rm D}Y_{\rm{el}}(F)$
&0&999840150   &0&333238563    
&0&499429548   &0&249553251
\\ \hline

$\De g_{\rm{int}}Y_{\rm{el}}(F)$
&0&000065379(10)  &0&000029443(5)
&0&00009031(2)    &0&000056806(18)
\\ \hline

$\De g_{\rm{QED}}Y_{\rm{el}}(F)$
&0&001159708(3)   &0&000386592(2)
&0&000580176(15)  &0&000290213(13)
\\ \hline

$\De g_{\rm{rec}}^{(e)}Y_{\rm{el}}(F)$
&0&000000005(1)  &0&000000003
&0&000000011(1)  &0&000000007 
\\ \hline

$\De g_{\rm{NS}}Y_{\rm{el}}(F)$
&0&0              &0&0     
&0&000000001      &0&000000002
\\ \hline

$-\frac{\me}{\mpr}g_{I}Y_{\rm{nuc}}^{(\mu)}(F)$
&-0&000382545(1)  &0&000343797(16) 
&-0&000175318     &0&000179403(1)
\\ \hline

$\delta g_{\rm{HFS}(\mu)}^{(2s)}$
&0&000000007(1)  &-0&000000010(1)
&0&000000011     &-0&000000015 
\\ \hline

$\delta g_{\rm{HFS}(Q)}^{(2s)}$
&\multicolumn{2}{c|}{------} &0&0
&0&0                         &0&0 
\\ \hline

$g$
&1&000682704(10)     &0&333998387(17)
&0&49992474(3)       &0&25007967(2) 
\\ \hline
\end{tabular}
\end{center}
\begin{center}
\begin{tabular}{||c|r@{.}l|r@{.}l|r@{.}l|r@{.}l||}
\hline
 Ion
&\multicolumn{2}{c|}{$^{53}\rm{Cr}^{21+}$}
&\multicolumn{2}{c|}{$^{73}\rm{Ge}^{29+}$}
&\multicolumn{2}{c|}{$^{129}\rm{Xe}^{51+}$}
&\multicolumn{2}{c||}{$^{131}\rm{Xe}^{51+}$}
\\ \hline
$I$ 
&\multicolumn{2}{c|}{3/2}
&\multicolumn{2}{c|}{9/2}
&\multicolumn{2}{c|}{1/2}
&\multicolumn{2}{c||}{3/2}
\\ \hline
$\mu/\mun$
&\multicolumn{2}{c|}{-0.47454(3)}
&\multicolumn{2}{c|}{-0.8794677(2)}
&\multicolumn{2}{c|}{-0.7779763(84)}
&\multicolumn{2}{c||}{0.6918619(39)}
\\ \hline
$Q$ (barn)
&\multicolumn{2}{c|}{-0.150(50)}
&\multicolumn{2}{c|}{-0.196}
&\multicolumn{2}{c|}{------}
&\multicolumn{2}{c||}{-0.114(1)}
\\ \hline
$g_{\rm D}Y_{\rm{el}}(F)$ 
&0&498709516 &0&199075231
&0&986375103 &0&493187551  
\\ \hline

$\De g_{\rm{int}}Y_{\rm{el}}(F)$
&0&00013712(5)  &0&00007397(4)     
&0&0006497(6) &0&0003249(3) 
\\ \hline

$\De g_{\rm{QED}}Y_{\rm{el}}(F)$
&0&00058076(4)   &0&00023270(2)
&0&0011736(2)  &0&00058679(11) 
\\ \hline

$\De g_{\rm{rec}}^{(e)}Y_{\rm{el}}(F)$
&0&000000017(1)  &0&000000009(1)
&0&00000008(3)   &0&000000040(15) 
\\ \hline

$\De g_{\rm{NS}}Y_{\rm{el}}(F)$
&0&000000009        &0&000000016 
&0&000001678(5)     &0&000000839(3)
\\ \hline

$-\frac{\me}{\mpr}g_{I}Y_{\rm{nuc}}^{(\mu)}(F)$
&0&000129221(8)   & 0&000095795 
&0&000423699(5)   &-0&000188400(1) 
\\ \hline

$\delta g_{\rm{HFS}(\mu)}^{(2s)}$
&-0&000000013  &-0&000000014 
&-0&000000130  & 0&000000058 
\\ \hline

$\delta g_{\rm{HFS}(Q)}^{(2s)}$
&0 &0                            &0&0 
&\multicolumn{2}{c|}{------}     &0&0 
\\ \hline

$g$
&0&49955663(7)     &0&19947771(4) 
&0&9886237(6)      &0&4939117(3)  
\\ \hline
\end{tabular}
\end{center}
\end{table}

\clearpage
\newpage
\begin{table}
Table 4 ({\it continued})
\begin{center}
\begin{tabular}{||c|r@{.}l|r@{.}l|r@{.}l||}
\hline
 Ion &\multicolumn{2}{c|}{$^{207}\rm{Pb}^{79+}$}&\multicolumn{2}{c|}{
$^{209}\rm{Bi}^{80+}$}
&\multicolumn{2}{c||}{$^{235}\rm{U}^{89+}$} \\ \hline
$I$
&\multicolumn{2}{c|}{1/2}
&\multicolumn{2}{c|}{9/2} 
&\multicolumn{2}{c||}{7/2}
\\ \hline
$\mu/\mun$
&\multicolumn{2}{c|}{0.59258(1)}
&\multicolumn{2}{c|}{4.1106(2)}
&\multicolumn{2}{c||}{-0.39(7)$^{a}$} \\ \hline
$Q$ (barn)
&\multicolumn{2}{c|}{------}
&\multicolumn{2}{c|}{-0.516(15)}
&\multicolumn{2}{c||}{4.936(6)}
\\ \hline
$g_{\rm D}Y_{\rm{el}}(F)$
&0&966001452
&0&193006882 
&0&238840328 
\\ \hline

$\De g_{\rm{int}}Y_{\rm{el}}(F)$
&0&0010703(14)  
&0&0002175(3)      
&0&0003127(5)  
\\ \hline

$\De g_{\rm{QED}}Y_{\rm{el}}(F)$
&0&0012023(7)
&0&00024077(13)  
&0&0003049(2)
\\ \hline

$\De g_{\rm{rec}}^{(e)}Y_{\rm{el}}(F)$
&0&00000013(18)
&0&00000003(4)
&0&00000004(9) 
\\ \hline

$\De g_{\rm{NS}}Y_{\rm{el}}(F)$
&0&00003921(8)
&0&000008752(17)
&0&00002999(6)
\\ \hline

$\De g_{\rm{NP}}Y_{\rm{el}}(F)$
&-0&00000002(2)
&\multicolumn{2}{c|}{------} 
&-0&00000003(3) 
\\ \hline

$-\frac{\me}{\mpr}g_{I}Y_{\rm{nuc}}^{(\mu)}(F)$
&-0&000322731(5)
&-0&00044774(2)
&0&000053(10)
\\ \hline

$\delta g_{\rm{HFS}(\mu)}^{(2s)}$
&0&000000253
&0&000000364 
&-0&000000062(11) 
\\ \hline

$\delta g_{\rm{HFS}(Q)}^{(2s)}$
&\multicolumn{2}{c|}{------}
&-0&000000001 
&0&000000021(1)
\\ \hline

$g$
&0&9679909(15)
&0&1930266(3) 
&0&239541(10)
\\ \hline
\end{tabular}
\end{center}
$^{a}$ An average of the values given in Ref. \cite{rag89}.
\end{table}

\clearpage
\newpage
\begin{figure}
\begin{picture}(450,120)
\multiput(0,0)(100,0){2}{\line(0,1){80}}
\put(0,55){\line(1,0){55}}
\put(55,55){\circle*{5}}
\multiput(0,25)(10,0){6}{\line(1,0){5}}
\put(55,25){\circle*{3}}
\multiput(100,55)(10,0){6}{\line(1,0){5}}
\put(155,55){\circle*{3}}
\put(100,25){\line(1,0){55}}
\put(155,25){\circle*{5}}
\put(80,40){+}
\put(55,65){$V_{\rm HFS}^{(\mu)}$}
\put(55,10){$V_{\vecb}^{(e)}$}
\put(155,65){$V_{\vecb}^{(e)}$}
\put(155,10){$V_{\rm HFS}^{(\mu)}$}
\put(80,100){$(\mu)$}
\multiput(250,0)(100,0){2}{\line(0,1){80}}
\put(250,55){\line(1,0){55}}
\put(305,55){\circle*{5}}
\multiput(250,25)(10,0){6}{\line(1,0){5}}
\put(305,25){\circle*{3}}
\multiput(350,55)(10,0){6}{\line(1,0){5}}
\put(405,55){\circle*{3}}
\put(350,25){\line(1,0){55}}
\put(405,25){\circle*{5}}
\put(330,40){+}
\put(305,65){$V_{\rm HFS}^{(Q)}$}
\put(305,10){$V_{\vecb}^{(e)}$}
\put(405,65){$V_{\vecb}^{(e)}$}
\put(405,10){$V_{\rm HFS}^{(Q)}$}
\put(330,100){$(Q)$}
\end{picture}
\caption{The second-order diagrams 
contributing to $\delta g_{\rm HFS}^{(2s)}$.}\label{Diagrams2} 
\end{figure}
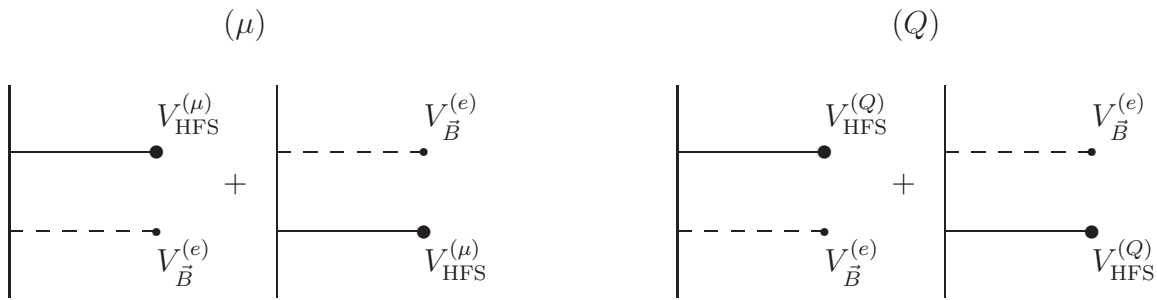
\begin{figure}
\begin{picture}(200,140)
\put(0,70){\vector(1,0){200}}
\put(100,0){\vector(0,1){140}}
\put(125,70){\circle{26}}
\put(125,70){\circle*{3}}
\put(125,84){\vector(-1,0){2}}
\put(105,49){$C_{\rm int}$}
\put(120,72){$1s$}
\put(0,60){\line(1,0){90}}
\put(0,40){\line(1,0){120}}
\put(108,40){\vector(1,0){2}}
\put(115,28){$C'$}
\put(90,60){\line(1,1){40}}
\put(120,40){\line(1,1){40}}
\put(130,100){\line(1,0){70}}
\put(160,80){\line(1,0){40}}
\put(180,100){\vector(1,0){2}}
\put(150,103){$C$}
\multiput(160,70)(5,0){2}{\circle*{3}}
\put(175,71){\line(1,0){22}}
\put(175,70.5){\line(1,0){22}}
\put(170,55){$+mc^2$}
\put(0,71){\line(1,0){22}}
\put(0,70.5){\line(1,0){22}}
\put(5,75){$-mc^2$}
\end{picture}
\caption{$C$ is the original contour of the integration
over the electron energy variable in the formalism with
the standard vacuum. $C'$ is the integration contour for 
the vacuum with the $(1s)^2$ shell included.
The integral along the contour $C_{\rm int}=C'-C$ describes 
the interaction of the valent electron with the $(1s)^2$-shell
electrons.}\label{E-plane} 
\end{figure}
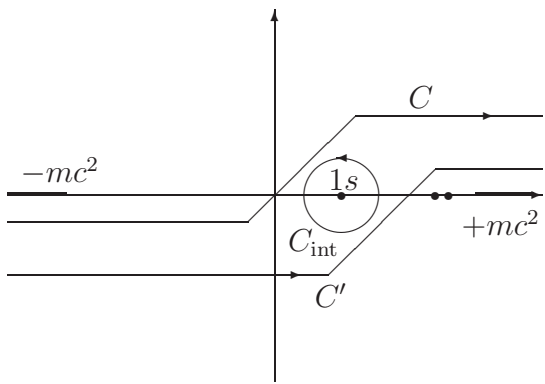
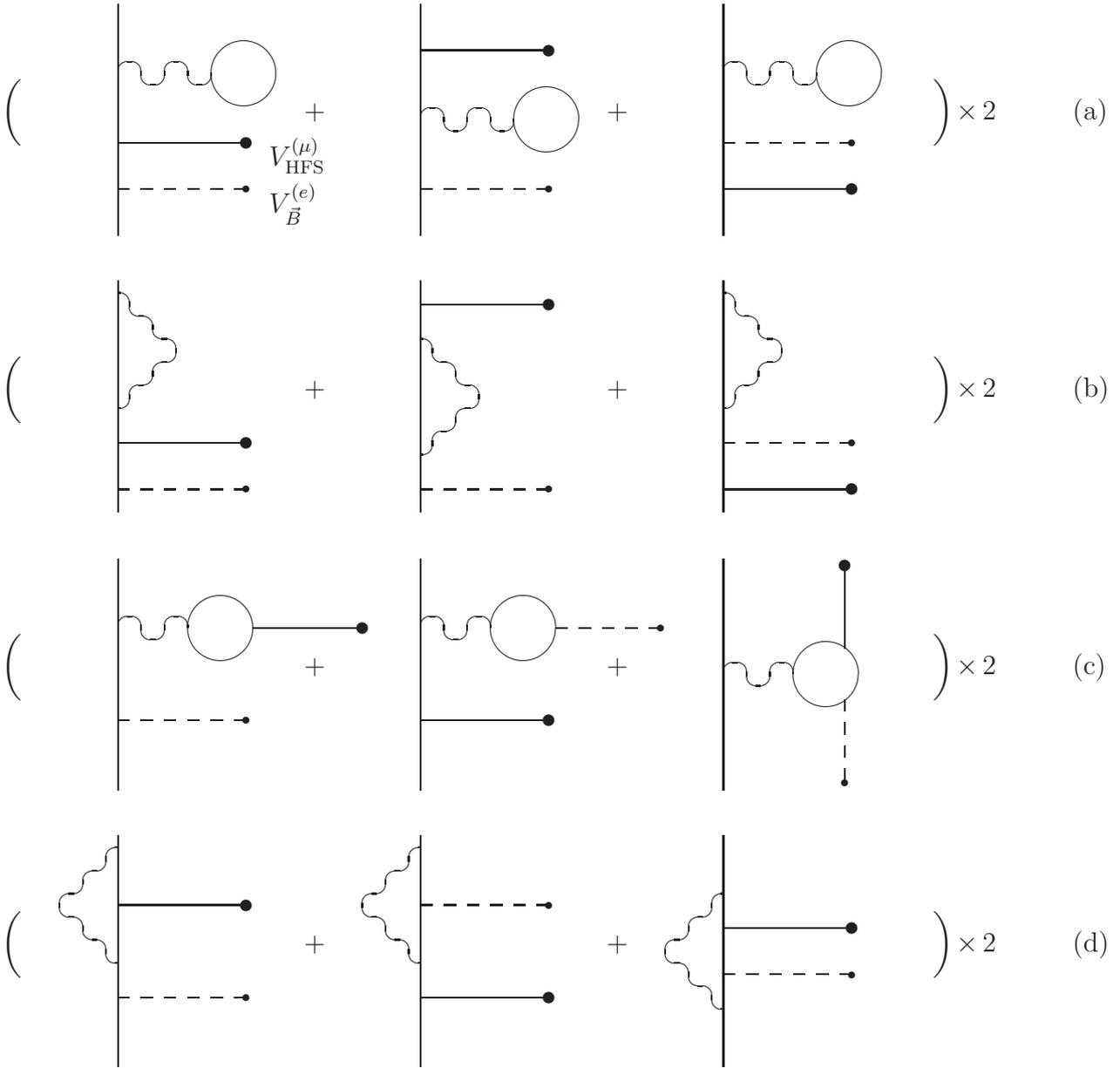
\begin{figure}
\begin{picture}(460,460)
\multiput(50,360)(130,0){3}{\line(0,1){100}}
\multiput(55,430)(20,0){2}{\oval(10,10)[t]}
\multiput(65,430)(20,0){2}{\oval(10,10)[b]}
\put(104,430){\circle{29}}
\put(50,400){\line(1,0){55}}
\put(105,400){\circle*{5}}
\multiput(50,380)(10,0){6}{\line(1,0){5}}
\put(105,380){\circle*{3}}
\put(180,440){\line(1,0){55}}
\put(235,440){\circle*{5}}
\multiput(185,410)(20,0){2}{\oval(10,10)[t]}
\multiput(195,410)(20,0){2}{\oval(10,10)[b]}
\put(234,410){\circle{29}}
\multiput(180,380)(10,0){6}{\line(1,0){5}}
\put(235,380){\circle*{3}}
\multiput(315,430)(20,0){2}{\oval(10,10)[t]}
\multiput(325,430)(20,0){2}{\oval(10,10)[b]}
\put(364,430){\circle{29}}
\multiput(310,400)(10,0){6}{\line(1,0){5}}
\put(365,400){\circle*{3}}
\put(310,380){\line(1,0){55}}
\put(365,380){\circle*{5}}
\put(0,410){$\biggl($}
\multiput(130,410)(130,0){2}{+}
\put(400,410){$\biggr)$}
\put(410,410){$\times\, 2$}
\put(115,390){$V_{\rm HFS}^{(\mu)}$}
\put(115,370){$V_{\vecb}^{(e)}$}
\put(460,410){(a)}
\multiput(50,240)(130,0){3}{\line(0,1){100}}
\put(50,330){\oval(10,10)[tr]}
\put(50,290){\oval(10,10)[br]}
\put(60,330){\oval(10,10)[bl]}
\put(60,290){\oval(10,10)[tl]}
\put(60,320){\oval(10,10)[tr]}
\put(60,300){\oval(10,10)[br]}
\put(70,320){\oval(10,10)[bl]}
\put(70,300){\oval(10,10)[tl]}
\put(70,310){\oval(10,10)[tr]}
\put(70,310){\oval(10,10)[br]}
\put(50,270){\line(1,0){55}}
\put(105,270){\circle*{5}}
\multiput(50,250)(10,0){6}{\line(1,0){5}}
\put(105,250){\circle*{3}}
\put(180,310){\oval(10,10)[tr]}
\put(180,270){\oval(10,10)[br]}
\put(190,310){\oval(10,10)[bl]}
\put(190,270){\oval(10,10)[tl]}
\put(190,300){\oval(10,10)[tr]}
\put(190,280){\oval(10,10)[br]}
\put(200,300){\oval(10,10)[bl]}
\put(200,280){\oval(10,10)[tl]}
\put(200,290){\oval(10,10)[tr]}
\put(200,290){\oval(10,10)[br]}
\put(180,330){\line(1,0){55}}
\put(235,330){\circle*{5}}
\multiput(180,250)(10,0){6}{\line(1,0){5}}
\put(235,250){\circle*{3}}
\put(310,330){\oval(10,10)[tr]}
\put(310,290){\oval(10,10)[br]}
\put(320,330){\oval(10,10)[bl]}
\put(320,290){\oval(10,10)[tl]}
\put(320,320){\oval(10,10)[tr]}
\put(320,300){\oval(10,10)[br]}
\put(330,320){\oval(10,10)[bl]}
\put(330,300){\oval(10,10)[tl]}
\put(330,310){\oval(10,10)[tr]}
\put(330,310){\oval(10,10)[br]}
\multiput(310,270)(10,0){6}{\line(1,0){5}}
\put(365,270){\circle*{3}}
\put(310,250){\line(1,0){55}}
\put(365,250){\circle*{5}}
\put(0,290){$\biggl($}
\multiput(130,290)(130,0){2}{+}
\put(400,290){$\biggr)$}
\put(410,290){$\times\, 2$}
\put(460,290){(b)}
\multiput(50,120)(130,0){3}{\line(0,1){100}}
\multiput(55,190)(20,0){2}{\oval(10,10)[t]}
\put(65,190){\oval(10,10)[b]}
\put(94,190){\circle{29}}
\put(108,190){\line(1,0){45}}
\put(155,190){\circle*{5}}
\multiput(50,150)(10,0){6}{\line(1,0){5}}
\put(105,150){\circle*{3}}
\multiput(185,190)(20,0){2}{\oval(10,10)[t]}
\put(195,190){\oval(10,10)[b]}
\put(224,190){\circle{29}}
\multiput(238,190)(10,0){5}{\line(1,0){5}}
\put(283,190){\circle*{3}}
\put(180,150){\line(1,0){55}}
\put(235,150){\circle*{5}}
\multiput(315,170)(20,0){2}{\oval(10,10)[t]}
\put(325,170){\oval(10,10)[b]}
\put(354,170){\circle{29}}
\put(362,181){\line(0,1){35}}
\put(362,217){\circle*{5}}
\multiput(362,159)(0,-10){4}{\line(0,-1){5}}
\put(362,123){\circle*{3}}
\put(0,170){$\biggl($}
\multiput(130,170)(130,0){2}{+}
\put(400,170){$\biggr)$}
\put(410,170){$\times\, 2$}
\put(460,170){(c)}
\multiput(50,0)(130,0){3}{\line(0,1){100}}
\put(50,90){\oval(10,10)[tl]}
\put(50,50){\oval(10,10)[bl]}
\put(40,90){\oval(10,10)[br]}
\put(40,50){\oval(10,10)[tr]}
\put(40,80){\oval(10,10)[tl]}
\put(40,60){\oval(10,10)[bl]}
\put(30,80){\oval(10,10)[br]}
\put(30,60){\oval(10,10)[tr]}
\put(30,70){\oval(10,10)[tl]}
\put(30,70){\oval(10,10)[bl]}
\put(50,70){\line(1,0){55}}
\put(105,70){\circle*{5}}
\multiput(50,30)(10,0){6}{\line(1,0){5}}
\put(105,30){\circle*{3}}
\put(180,90){\oval(10,10)[tl]}
\put(180,50){\oval(10,10)[bl]}
\put(170,90){\oval(10,10)[br]}
\put(170,50){\oval(10,10)[tr]}
\put(170,80){\oval(10,10)[tl]}
\put(170,60){\oval(10,10)[bl]}
\put(160,80){\oval(10,10)[br]}
\put(160,60){\oval(10,10)[tr]}
\put(160,70){\oval(10,10)[tl]}
\put(160,70){\oval(10,10)[bl]}
\multiput(180,70)(10,0){6}{\line(1,0){5}}
\put(235,70){\circle*{3}}
\put(180,30){\line(1,0){55}}
\put(235,30){\circle*{5}}
\put(310,70){\oval(10,10)[tl]}
\put(310,30){\oval(10,10)[bl]}
\put(300,70){\oval(10,10)[br]}
\put(300,30){\oval(10,10)[tr]}
\put(300,60){\oval(10,10)[tl]}
\put(300,40){\oval(10,10)[bl]}
\put(290,60){\oval(10,10)[br]}
\put(290,40){\oval(10,10)[tr]}
\put(290,50){\oval(10,10)[tl]}
\put(290,50){\oval(10,10)[bl]}
\put(310,60){\line(1,0){55}}
\put(365,60){\circle*{5}}
\multiput(310,40)(10,0){6}{\line(1,0){5}}
\put(365,40){\circle*{3}}
\put(0,50){$\biggl($}
\multiput(130,50)(130,0){2}{+}
\put(400,50){$\biggr)$}
\put(410,50){$\times\, 2$}
\put(460,50){(d)}
\end{picture}
\caption{The third-order diagrams 
contributing to $\delta g_{\rm HFS(\mu)}^{(2s)}$.}
\label{Diagrams3} 
\end{figure}
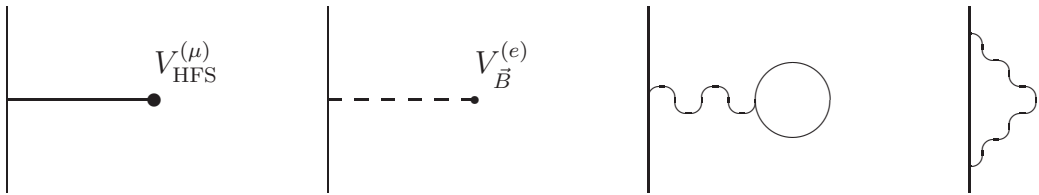
\begin{figure}
\begin{picture}(460,70)
\multiput(50,0)(120,0){4}{\line(0,1){70}}
\put(50,35){\line(1,0){55}}
\put(105,35){\circle*{5}}
\put(105,45){$V_{\rm HFS}^{(\mu)}$}
\multiput(170,35)(10,0){6}{\line(1,0){5}}
\put(225,35){\circle*{3}}
\put(225,45){$V_{\vecb}^{(e)}$}
\multiput(295,35)(20,0){2}{\oval(10,10)[t]}
\multiput(305,35)(20,0){2}{\oval(10,10)[b]}
\put(344,35){\circle{29}}
\put(410,55){\oval(10,10)[tr]}
\put(410,15){\oval(10,10)[br]}
\put(420,55){\oval(10,10)[bl]}
\put(420,15){\oval(10,10)[tl]}
\put(420,45){\oval(10,10)[tr]}
\put(420,25){\oval(10,10)[br]}
\put(430,45){\oval(10,10)[bl]}
\put(430,25){\oval(10,10)[tl]}
\put(430,35){\oval(10,10)[tr]}
\put(430,35){\oval(10,10)[br]}
\end{picture}
\caption{The first-order diagrams 
contributing to $\delta g_{\rm HFS(\mu)}^{(2s)}$ being multiplied
by the second-order diagrams 
presented in Fig. \ref{Diagrams32}.}\label{Diagrams31} 
\end{figure}
\begin{figure}
\begin{picture}(460,280)
\multiput(50,200)(130,0){3}{\line(0,1){80}}
\put(50,255){\line(1,0){55}}
\put(105,255){\circle*{5}}
\multiput(50,225)(10,0){6}{\line(1,0){5}}
\put(105,225){\circle*{3}}
\put(25,240){$2\,\times$}
\put(115,250){$V_{\rm HFS}^{(\mu)}$}
\put(115,220){$V_{\vecb}^{(e)}$}
\multiput(185,255)(20,0){2}{\oval(10,10)[t]}
\multiput(195,255)(20,0){2}{\oval(10,10)[b]}
\put(234,255){\circle{29}}
\put(180,225){\line(1,0){55}}
\put(235,225){\circle*{5}}
\put(155,240){$2\,\times$}
\multiput(315,255)(20,0){2}{\oval(10,10)[t]}
\multiput(325,255)(20,0){2}{\oval(10,10)[b]}
\put(364,255){\circle{29}}
\multiput(310,225)(10,0){6}{\line(1,0){5}}
\put(365,225){\circle*{3}}
\put(285,240){$2\,\times$}
\multiput(50,100)(130,0){3}{\line(0,1){80}}
\put(50,170){\oval(10,10)[tr]}
\put(50,130){\oval(10,10)[br]}
\put(60,170){\oval(10,10)[bl]}
\put(60,130){\oval(10,10)[tl]}
\put(60,160){\oval(10,10)[tr]}
\put(60,140){\oval(10,10)[br]}
\put(70,160){\oval(10,10)[bl]}
\put(70,140){\oval(10,10)[tl]}
\put(70,150){\oval(10,10)[tr]}
\put(70,150){\oval(10,10)[br]}
\put(50,112){\line(1,0){55}}
\put(105,112){\circle*{5}}
\put(25,140){$2\,\times$}
\put(180,170){\oval(10,10)[tr]}
\put(180,130){\oval(10,10)[br]}
\put(190,170){\oval(10,10)[bl]}
\put(190,130){\oval(10,10)[tl]}
\put(190,160){\oval(10,10)[tr]}
\put(190,140){\oval(10,10)[br]}
\put(200,160){\oval(10,10)[bl]}
\put(200,140){\oval(10,10)[tl]}
\put(200,150){\oval(10,10)[tr]}
\put(200,150){\oval(10,10)[br]}
\multiput(180,112)(10,0){6}{\line(1,0){5}}
\put(235,112){\circle*{3}}
\put(155,140){$2\,\times$}
\multiput(315,140)(20,0){2}{\oval(10,10)[t]}
\put(325,140){\oval(10,10)[b]}
\put(354,140){\circle{29}}
\put(368,140){\line(1,0){45}}
\put(415,140){\circle*{5}}
\multiput(50,0)(130,0){3}{\line(0,1){80}}
\put(50,60){\oval(10,10)[tl]}
\put(50,20){\oval(10,10)[bl]}
\put(40,60){\oval(10,10)[br]}
\put(40,20){\oval(10,10)[tr]}
\put(40,50){\oval(10,10)[tl]}
\put(40,30){\oval(10,10)[bl]}
\put(30,50){\oval(10,10)[br]}
\put(30,30){\oval(10,10)[tr]}
\put(30,40){\oval(10,10)[tl]}
\put(30,40){\oval(10,10)[bl]}
\put(50,40){\line(1,0){55}}
\put(105,40){\circle*{5}}
\put(180,60){\oval(10,10)[tl]}
\put(180,20){\oval(10,10)[bl]}
\put(170,60){\oval(10,10)[br]}
\put(170,20){\oval(10,10)[tr]}
\put(170,50){\oval(10,10)[tl]}
\put(170,30){\oval(10,10)[bl]}
\put(160,50){\oval(10,10)[br]}
\put(160,30){\oval(10,10)[tr]}
\put(160,40){\oval(10,10)[tl]}
\put(160,40){\oval(10,10)[bl]}
\multiput(180,40)(10,0){6}{\line(1,0){5}}
\put(235,40){\circle*{3}}
\multiput(315,40)(20,0){2}{\oval(10,10)[t]}
\put(325,40){\oval(10,10)[b]}
\put(354,40){\circle{29}}
\multiput(368,40)(10,0){5}{\line(1,0){5}}
\put(415,40){\circle*{3}}
\end{picture}
\caption{The second-order diagrams 
contributing to $\delta g_{\rm HFS(\mu)}^{(2s)}$ 
being multiplied by the 
first-order diagrams presented 
in Fig. \ref{Diagrams31}.}\label{Diagrams32} 
\end{figure}
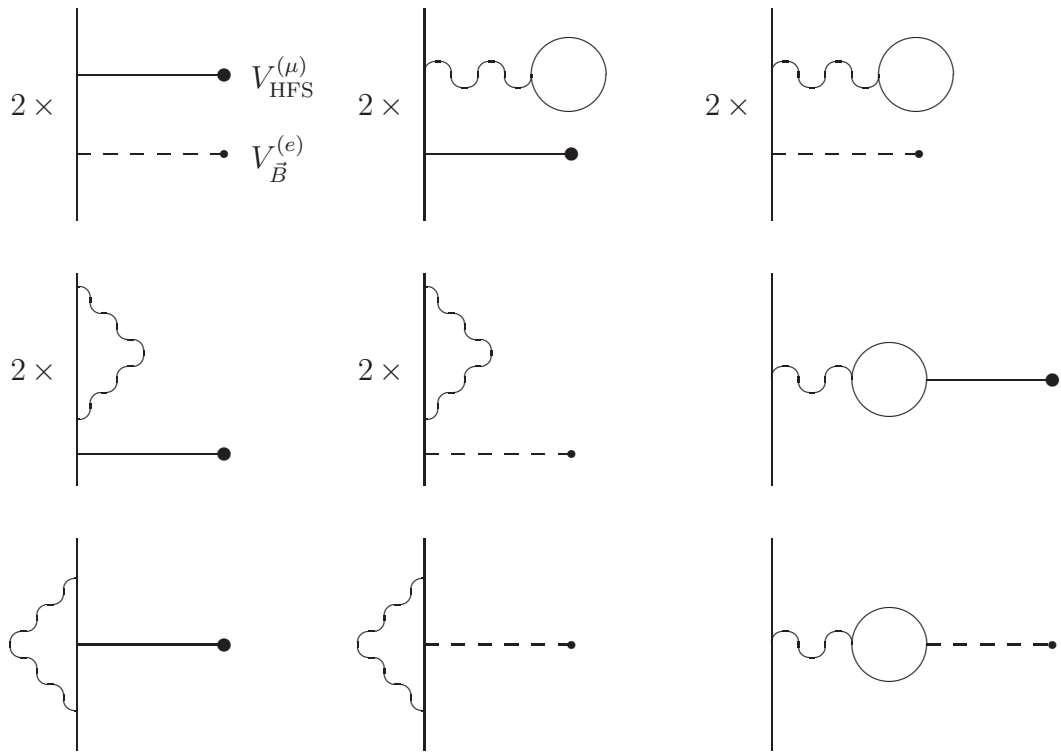

\end{document}